# Phase transitions in a frustrated biquadratic Heisenberg model with coupled orbital degree of freedom for iron-based superconductors

W. Z. Zhuo[1], M. H. Qin[1 a)], S. Dong[2], X. G. Li[3], and J. –M. Liu[1, 4 b)]

[1]*Institute for Advanced Materials and Guangdong Provincial Key Laboratory of Quantum Engineering and Quantum Materials, South China Normal University, Guangzhou 510006, China*

[2]*Department of Physics, Southeast University, Nanjing 211189, China*

[3]*Hefei National Laboratory for Physical Sciences at Microscale, Department of Physics, University of Science and Technology of China, Hefei 230026, China*

[4]*Laboratory of Solid State Microstructures, Nanjing University, Nanjing 210093, China*

**[Abstract]** In this work, we study a biquadratic Heisenberg model with coupled orbital degree of freedom using Monte Carlo simulation in order to investigate the phase transitions in iron-based superconductors. The antiferro-quadrupolar state, which may be related to the magnetism of FeSe [Phys. Rev. Lett. 115, 116401 (2015)], is stabilized by the anisotropic biquadratic interaction induced by a ferro-orbital-ordered state. It is revealed that the orbital and nematic transitions occur at the same temperature for all the cases, supporting the mechanism of the orbital-driven nematicity as revealed in most recent experiments [Nat. Mater. 14, 210 (2015)]. In addition, it is suggested that the orbital interaction may lead to the separation of the structural and magnetic phase transitions as observed in many families of iron pnictides.

---

* Electronic mail: a) qinmh@scnu.edu.cn and b) liujm@nju.edu.cn

**I. Introduction**

In the past few years, fascinating structural and magnetic phase transitions in iron-based superconductors have drawn extensive attention both experimentally and theoretically.[1-5] Experimentally, the collinear ($\pi$, 0) antiferromagnetic (AFM) order is developed at low temperatures (*T*) in most iron pnictides, and usually accompanied by a tetragonal-to-orthorhombic distortion.[6-8] The structural transition temperature $T_S$ is either equal to or higher than the AFM transition temperature $T_{AFM}$: $T_S \geq T_{AFM}$. In order to understand this phenomenon, two different mechanisms emphasizing the essential role of spin[9,10] and orbital[11,12] fluctuations have been proposed, respectively. Specifically, several theoretical calculations on models for pnictides suggest that the nematic order can be developed before the stabilization of the AFM order, and drives the structural phase transition.[13-16] The experimental observations of the spin excitation spectrum can be well explained based on this mechanism.[17,18] However, the magnetism-based origin is strongly challenged by the recent experiments in FeSe.[19,20]

As one of the most famous iron chalcogenide superconductors, FeSe shows a very high $T_C$ superconductivity in its single-layer limit.[21-23] Interestingly, lowing *T* in bulk FeSe leads to a tetragonal-to-orthorhombic structural transition, while no long-range AFM transition has been observed, suggesting that the structural transition of FeSe may not have the magnetic origin.[24] Most recently, a theoretical study considering only the localized spins revealed a nematic quantum paramagnetic phase caused by quantum fluctuations and strongly frustrated exchange interactions, which may contribute to the nematicity in FeSe.[25] In addition, a frustrated bilinear-biquadratic Heisenberg model describing the magnetism of FeSe was studied, and the antiferroquadrupolar (AFQ) and Ising-nematic orders were identified at low *T*.[26] The structural phase transition in FeSe was suggested to correspond to the Ising-nematic transition in the model. Furthermore, it was proposed that the Goldstone modes of the AFQ order may contribute to the low-energy dipolar magnetic fluctuations observed in the nuclear magnetic resonance measurements.[24,27] In addition, an unusual magnetic frustration was proposed in a theoretical work and suggested to suppress magnetic order and trigger ferro-orbital order in the nematic phase, consistent with the pressure dependence of $T_C$ in FeSe.[28]

On the other hand, recent experiments clearly demonstrated that the orbital transition may play a crucial role in the onset of nematic transition in bulk FeSe, strongly supporting the mechanism of orbital-driven nematicity.[27] In addition, it was observed that the shear-modulus softening above $T_S$ in FeSe and underdoped BaFe$_2$As$_2$ are nearly identical, suggesting a common origin for the structural transition in these materials.[24] As a matter of fact, the important role of orbital fluctuations has been addressed in several models for pnictides.[29-32] For example, the nature of phase transitions in pnictides has been investigated based on a spin-orbital model with coupled spin and orbital degrees of freedom using Monte Carlo (MC) simulations.[33] However, the biquadratic interaction which is believed to be very important in iron superconductors[34] is neglected in this model, while a coupling to orbital degree of freedom is not considered in the localized spin models describing the magnetism of FeSe discussed above. Thus, more model calculations studying the role of orbital fluctuations in iron pnictides and chalcogenides are still urgently needed in order to elucidate the physical origins for the nematicity and some other experimental observations.

In this work, we study a classical biquadratic Heisenberg model with coupled orbital degree of freedom. Several experimental observations in iron-based superconductors can be qualitatively explained in our MC simulations. In detail, both the AFM and AFQ orders are predicted in the phase diagram of the model. The orbital phase transition temperature $T_O$ is always the same as or higher than the AFM (AFQ) transition temperature $T_{AFM}$ ($T_{AFQ}$), and a nematic ordering can also be developed at $T^* \approx T_O$, supporting the mechanism of orbital-driven nematicity. The remainder of this paper is organized as follows: in Sec. II the model and the simulation method will be described. Sec. III is attributed to the simulation results and discussion, and the conclusion is presented in Sec. IV.

## II. Model and method

We study a classical biquadratic Heisenberg spin ($S = 1$) model with coupled spin and orbital degrees of freedom on a two-dimensional square lattice. It was reported that a ferro-orbital order in which the $d_{xz}/d_{yz}$ orbital dominates can be developed in iron pnictides and FeSe as $T$ decreases.[18,27,35] For simplicity, the orbital parameter $O$ is set to be Ising variables with values 1 (occupation of $d_{xz}$ orbital) or $-1$ (occupation of $d_{yz}$ orbital). The model

Hamiltonian is stated as:

$$H = \sum_i (J_1 + \Delta_i^x \cdot J_{OS}) \vec{S}_i \cdot \vec{S}_{i+x} + \sum_i (J_1 + \Delta_i^y \cdot J_{OS}) \vec{S}_i \cdot \vec{S}_{i+y} + J_2 \sum_{[ij]} \vec{S}_i \cdot \vec{S}_j$$
$$- \sum_i (K_1 + \Delta_i^x \cdot K_{OS})(\vec{S}_i \cdot \vec{S}_{i+x})^2 - \sum_i (K_1 + \Delta_i^y \cdot K_{OS})(\vec{S}_i \cdot \vec{S}_{i+y})^2 - J_O \sum_{\langle ij \rangle} O_i \cdot O_j, \quad (1)$$

with

$$\Delta_i^x = \begin{cases} 1 & O_i = O_{i+x} = 1 \\ 0 & O_i = -1 \text{ or } O_{i+x} = -1 \end{cases}$$
$$\Delta_i^y = \begin{cases} 1 & O_i = O_{i+y} = -1 \\ 0 & O_i = 1 \text{ or } O_{i+y} = 1 \end{cases}, \quad (2)$$

The first two terms are the nearest neighbor (NN) exchange interactions alone the *x* and *y* directions. Following earlier work, the spin and orbital degrees of freedom are considered to be coupled by a Kugel-Khomskii-like mechanism[33,36] with the coupling magnitude $J_{OS}$. The third term denotes the next nearest neighbor exchange interaction with coupling $J_2$. The fourth and fifth terms are the NN biquadratic interactions alone the *x* and *y* directions, respectively, which are expected in any model calculations for iron superconductors.[34] In fact, it has been proved that the biquadratic interaction is also affected by the Fe-Fe bond length[37] which can be modulated by orbital orderings. Thus, the biquadratic interactions should be also dependent on orbital orders (with coupling $K_{OS}$). At last, the orbital interaction with coupling $J_O$ is considered, which has been involved in several earlier models.[12]

Generally, the introduce of the coupled orbital degree of freedom and biquadratic interaction in the Hamiltonian can be interpreted to change the linear-response exchange constant, which can be defined by:[13]

$$J_{ij} \equiv -(\vec{S}_i \cdot \vec{S}_j)^{-1} \frac{\partial^2 H}{\partial \theta^2}, \quad (3)$$

where $\theta$ is the angle between spins at sites *i* and *j*. For the ($\pi$, 0) AFM ground state normally accompanied by an $d_{xz}$ orbital polarization, as experimentally reported in several iron pnictides,[35] we get $J_{ij}^x = J_1 + \Delta_i^x(J_{OS} + 2K_{OS}) + 2K_1$. Furthermore, it is suggested that the magnitude of $J_{ij}^x$ may be increased when the sites *i* and *j* are occupied by $d_{xz}$ orbital.[12] Thus, we choose $K_{OS} > 0$ and $J_{OS} > 0$ in the whole work, and set $K_{OS} = 1$ as the energy unit, for simplicity. In addition, we fix $J_2 = 0.7J_1$ and $J_{OS} = 0.5J_1$, which are comparable with experimental reports.[38] Furthermore, we take the other coupling parameters as variables and study the phase diagram of the model using the standard Metropolis algorithm and temperature exchange method.[39,40] Unless stated elsewhere, the simulation is performed on a $36 \times 36$ lattice with periodic boundary conditions. To study the finite $T$ phase transitions to long-range magnetic orders, we introduce a spin-space anisotropy:

$$\vec{S}_i \cdot \vec{S}_j = S_i^z S_j^z + \lambda \left( S_i^x S_j^x + S_i^y S_j^y \right), \tag{4}$$

In order to explore the phases in the system, the dipolar and quadrupolar magnetic structure factors are calculated by:[26]

$$\Phi_d(\vec{q}) = \frac{1}{N^2} \sum_{ij} \left\langle \vec{S}_i \cdot \vec{S}_j \right\rangle \exp\left[ i\vec{q} \cdot (\vec{r}_i - \vec{r}_j) \right], \tag{5}$$

and

$$\Phi_q(\vec{q}) = \frac{1}{N^2} \sum_{ij} \left\langle \vec{Q}_i \cdot \vec{Q}_j \right\rangle \exp\left[ i\vec{q} \cdot (\vec{r}_i - \vec{r}_j) \right], \tag{6}$$

respectively. Here, $N$ is the number of sites, and $\langle \ldots \rangle$ is the ensemble average. For classical Heisenberg spins:

$$\vec{Q}_i \cdot \vec{Q}_j = 2\left( \vec{S}_i \cdot \vec{S}_j \right)^2 - \frac{2}{3} \vec{S}_i^{\,2} \cdot \vec{S}_j^{\,2}, \tag{7}$$

Then, the AFM and AFQ order parameters ($m$ and $m_q$) and the corresponding nematic order parameters ($N_{AFM}$ and $N_{AFQ}$) are calculated by:[13,26]

$$\begin{aligned} m^2 &= \Phi_d(\pi, 0) + \Phi_d(0, \pi) \\ m_q^2 &= \Phi_q(\pi, 0) + \Phi_q(0, \pi) \end{aligned}, \tag{8}$$

$$N_{AFM} = \left\langle \left\| \frac{1}{2N} \sum_i^N \left( \vec{S}_i \cdot \vec{S}_{i+x} - \vec{S}_i \cdot \vec{S}_{i+y} \right) \right\| \right\rangle$$

$$N_{AFQ} = \left\langle \left\| \frac{1}{2N} \sum_i^N \left( \vec{Q}_i \cdot \vec{Q}_{i+x} - \vec{Q}_i \cdot \vec{Q}_{i+y} \right) \right\| \right\rangle$$

(9)

The transition temperatures ($T_O$, $T^*$, $T_{AFQ}$, and $T_{AFM}$) can be estimated from the positions of the peaks in the corresponding susceptibility $\chi_O$, $\chi_N$, $\chi_{AFQ}$, and $\chi_{AFM}$, respectively.

### III. Simulation results and discussion

*A. The AFM and AFQ orders*

First, we study the spin orders developed at low $T$ for various $J_1$ and $K_1$. Fig. 1 shows the simulated phase diagram in the ($J_1$, $K_1$) plane at $T = 0.01$ for $\lambda = 0.95$ and $J_O = 0$. The phase diagram is occupied separately by the AFM order and AFQ order. It is indicated that the strong competition between the exchange interaction and the biquadratic interaction can stabilize the AFQ order. Here, it should be noted that in the bilinear-biquadratic spin model, an isotropic NN biquadratic interaction alone cannot stabilize the AFQ order.[26] In order to understand the physics underlying our simulated results on the AFQ order, a qualitative discussion in the energy landscape is helpful.

Fig. 2(a) and 2(b) show the distribution of dipolar and quadrupolar magnetic structure factors at $T = 0.01$ for $J_1 = 0.1$, $K_1 = -0.5$, $J_O = 0$ and $\lambda = 0.6$, as an example. A dominant ($\pi$, 0) AFQ correlation is clearly shown, with the three spin components of the AFQ order plotted in Fig. 2(c) and 2(d), respectively. For an AFQ order, the NN spins along the $x$ direction are perpendicular to each other, while those along the $y$ direction are antiparallel with each other. In this case, the $d_{yz}$ ferro-orbital-ordered state is developed at $T_O$ (not shown here), resulting in the anisotropic NN biquadratic interaction ($K_1$ along the $x$ direction, and $K_1 + 1$ along the $y$ direction). Thus, the NN spins along the $x$ direction tend to be perpendicular to each other to save the negative biquadratic interaction $K_1$ at the cost of the AFM exchange interaction $J_1$, while those along the $y$ direction are antiparallel with each other to satisfy the AFM exchange and positive biquadratic interactions, leading to the AFQ state. With increasing $|K_1|$ ($K_1 < 0$), the AFQ state can be further stabilized, resulting in the enlargement of the $J_1$-region with the AFQ order. On the other hand, the collinear AFM order can be stabilized by a positive biquadratic interaction ($K_1 > 0$), and a positive $K_1$ never stabilizes the AFQ state, as confirmed in our simulations.

B. The AFQ phase transitions

Given the AFQ order in the phase diagram, we subsequently draw our attention onto the expected nematic order and the finite-$T$ phase transitions. The order parameters for these phase transitions include orbital parameter $O_z = \langle \Sigma O_i \rangle / N$, $N_{AFQ}$, and $m_q$. Fig. 3(a) shows these parameters as a function of $T$ for $J_1 = 0.1$, $K_1 = -0.5$, $J_O = 0$ and $\lambda = 0.6$ as an example. At high $T$, the system is in the paramagnetic (PM) state, and all these order parameters are rather small. When $T$ falls down to the critical points, these parameters increase from the baseline, fingering the development of orbital order, nematic order, and AFQ order, respectively. These transition points ($T_O$, $T^*$, and $T_{AFQ}$) are estimated from the positions of peaks in the calculated susceptibility on these parameters, as shown in Fig. 3(b). It is clearly shown that the nematic order is developed at the same temperature as the orbital order does, giving $T^* \sim T_O$, consistent with experimental observations.[27] Furthermore, the AFQ order is stabilized at a slightly lower $T$, giving $T_{AFQ} < T_O$. Similar behaviors are confirmed for different lattice sizes and the conclusion regarding the nematic state is reliable.

The effects of spin anisotropy parameter $\lambda$, orbital interaction $J_O$ (i.e. ferro-orbital coupling), and biquadratic interaction $K_1$ on the phase transitions are also investigated, respectively, and the simulated phase diagrams are plotted in Fig. 4. Fig. 4(a) shows the phase diagram on the ($\lambda$, $T$) space for $J_1 = 0.1$, $K_1 = -0.5$ and $J_O = 0$. A reduced $\lambda$ (increased spin anisotropy) downshifts all the three transitions points. This is understandable since the spin anisotropy favors the AFM order rather than the AFQ order. At the same time, the coupling between the spin and orbital degrees of freedom in turn destabilizes the orbital/nematic order. The nematic order in advantage of the AFQ order can be destroyed eventually by the spin anisotropy, leading to narrowed $|T_{AFQ} - T^*|$ with increasing $(1-\lambda)$, i.e. the nematic order region is suppressed.

On the other hand, the orbital order can be further stabilized by a ferro-orbital coupling $J_O$, as shown in Fig. 4(b) for the phase diagram on the ($J_O$, $T$) parameter space for $J_1 = 0.1$, $K_1 = -0.5$, and $\lambda = 0.4$. Interestingly, the nematic order region is substantially enlarged as $J_O$ increases, and $T^* \sim T_O$ applies for all these cases, again demonstrating the important role of the orbital order in driving the nematicity in FeSe. Furthermore, Fig. 4(c) shows the phase diagram on the ($K_1$, $T$) parameter space for $J_1 = 0.1$, $Jo = 0$, and $\lambda = 0.6$. The three transition temperatures decrease as the magnitude of $K_1$ increases. It is noted that the energy gain from the biquadratic interaction due to the phase transition from the PM state to the AFQ state is extensively decreased with the increasing $|K_1|$, leading to the destabilization of the AFQ state. In turn, the nematic state and the orbital order are also destabilized, as shown in our simulations.

*C. The AFM phase transitions*

For integrity, we also study the collinear AFM phase transition for the cases of dominant $J_1$. The order parameters and their susceptibilities at $T = 0.01$ for $J_1 = 1$, $K_1 = 0.5$, $J_O = 0$, and $\lambda = 0.95$ are given in Fig. 5(a) and 5(b), respectively, clearly showing that the three phase transitions occur at the same temperature, i.e. $T_{AFM} \sim T^* \sim T_O$. In fact, this phenomenon was reported in some iron pnictides such as undoped 122 compounds $CaFe_2As_2$ and $SrFe_2As_2$.[41,42] Most recently, the ferro-orbital order has been proved to generate the exchange anisotropy and stabilize the collinear AFM order in iron-based superconductors by spin-wave analysis of the spin-fermion model,[43,44] well consistent with our simulations.[45]

Similarly, the phase transitions can be also modulated by spin anisotropy, orbital interaction, and biquadratic interaction, and the corresponding results are given in Fig. 6. As discussed earlier, a collinear AFM order is favored by the spin anisotropy, and $T_{AFM}$ will increase with decreasing $\lambda$, as confirmed in Fig. 6(a) which gives the phase diagram on the ($\lambda$, $T$) space for $J_1 = 1.0$, $K_1 = 0.5$, and $J_O = 0.2$. Furthermore, the nematic order region is also suppressed with increasing $(1-\lambda)$, and all the three transitions occur at the same temperature for $\lambda < 0.8$. Fig. 6(b) shows the phase diagram on the ($J_O$, $T$) space for $J_1 = 1.0$, $K_1 = 0.5$, and $\lambda = 0.95$. All the three transition temperatures increase with increasing $J_O$, and a separation between $T_{AFQ}$ and $T^*$ can be observed for $J_O > 0.1$. The phase diagram on the ($K_1$, $T$) space for $J_1 = 1.0$, $\lambda = 0.95$, and $J_O = 0.2$ is shown in Fig. 6(c). All the three transitions shift toward high $T$ as $K_1$ increases due to the increasing positive biquadratic interaction.

The nature of these transitions is also investigated by the analysis of the Binder ratios. Fig.

7 shows the Binder ratios for orbital $g_O$, nematic $g_N$ and spin $g_S$ ($g_O = \langle O_z^4 \rangle/\langle O_z^2 \rangle^2$, as an example) vs $T$ for $J_1 = 1.0$, $K_1 = 0.5$, and $\lambda = 0.95$. The clear divergences in the three curves for $J_O = 0$ indicate that these transitions are of the first-order. Interestingly, the divergences of $g_O$ and $g_N$ vanish when $J_O$ is increased above 0.1, as shown in Fig. 7(b), suggesting a trend to the second order orbital and nematic transitions. In addition, there is still a weak divergence in the spin Binder ratio $g_S$, which is indicative of a weak/pseudo-first order transition.[46] The nature of the AFM transition should be further checked by simulations on larger lattice sizes.[47] However, it is observed that the nematic transition always occurs at the same temperature as that of the ferro-orbital ordering, suggesting that the same mechanism for nematicity may also work in iron pnictides, in some extent.

*D. Relevance to iron superconductors*

Most recently, it was experimentally reported that spin-lattice relaxation rates are not affected at the nematic transition point, strongly suggesting the orbital origin for the nematicity in FeSe.[27] This viewpoint is supported in our simulations in which orbital order is suggested to be the primary driving mechanism for the finite temperature transitions. Furthermore, the AFQ order has been observed in the bilinear-biquadratic model and its Goldstone modes are suggested to contribute to dipolar magnetic fluctuations (in the absence of any AFM order) revealed in experiments.[26] However, a rather strong biquadratic interaction between the next nearest neighbors, which has not been confirmed in first-principles calculations,[28] is proved to be necessary to stabilize the AFQ order. In this work, it is suggested that the AFQ order can be stabilized by an anisotropic NN biquadratic interaction induced by the development of orbital order. Furthermore, the AFQ state can be replace by the AFM state by tuning the exchange and biquadratic interactions, as shown in the phase diagram in Fig. 1, which may contribute to the experimental observations of the appearance of magnetic order under external pressure in FeSe.[48,49] In addition, the nematic order accompanied by the orbital order may be developed in advance of the AFQ order when the orbital-orbital interaction is increased, further demonstrating the important role of the orbital fluctuations in the phase transitions in FeSe.

On the other hand, in iron pnictides $CaFe_2As_2$ and $SrFe_2As_2$, both the structural and magnetic phase transitions occurs at a same temperature, which is reproduced in our simulations for small $J_O < 0.1$. As a matter of fact, the phase transition in iron pncitides have been studied based on a similar spin-orbital model, and several behaviors are captured by the simulated phase diagram. For example, even for very weak spin anisotropy ($\lambda \sim 0.95$), both the orbital and AFM transition points coincide with each other, consistent with our simulations for $J_O = 0$. Interestingly, our work suggests that a ferro-orbital interaction may separate the two transitions and modulate the nature of the transitions. This behavior may contribute to some of the experimental observations in many families of iron pnictides. For example, both $T_S$ and $T_{AFM}$ in $BaFe_2As_2$ are increased, and $T_S$ becomes well above $T_{AFM}$ when a compressive stress is applied.[50,51] In some extent, this phenomenon may be related to the increase of the orbital-orbital interaction, which deserves to be checked further.

## IV. Conclusion

In conclusion, we have studied phase transitions in a biquadratic Heisenberg model with coupled orbital degree of freedom. It is suggested that the development of the ferro-orbital-ordered state induces the anisotropic biquadratic interaction between nearest neighbors, and in turn stabilizes the antiferroquadrupolar state which may be related to the magnetism of FeSe. For all the cases, the orbital and nematic transitions occur at a same temperature, supporting the mechanism of the orbital-driven nematicity as revealed in experiments. Furthermore, the orbital-orbital interaction may contribute to the separation of the structural and magnetic phase transitions as observed in many families of iron pnictides.

**Acknowledgements**:

The work is supported by the Natural Science Foundation of China (51332007, 51322206, 11274094), and the National Key Projects for Basic Research of China (2015CB921202 and 2015CB654602).

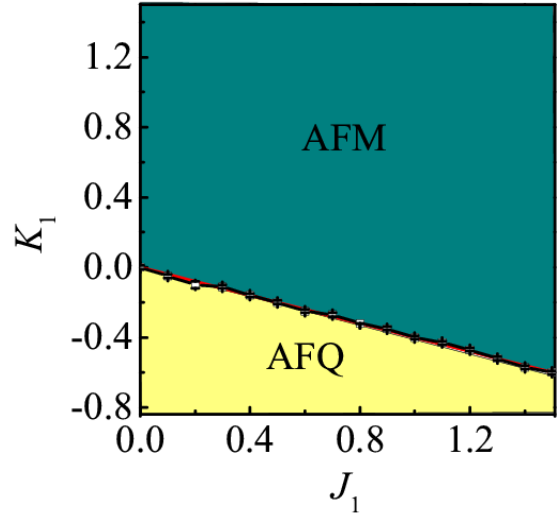

Figure 1. (color online) Calculated phase diagram in the $(J_1, K_1)$ parameter space at $T = 0.01$ for $\lambda = 0.95$ and $J_O = 0$. The red line shows the boundary at $T = 0$.

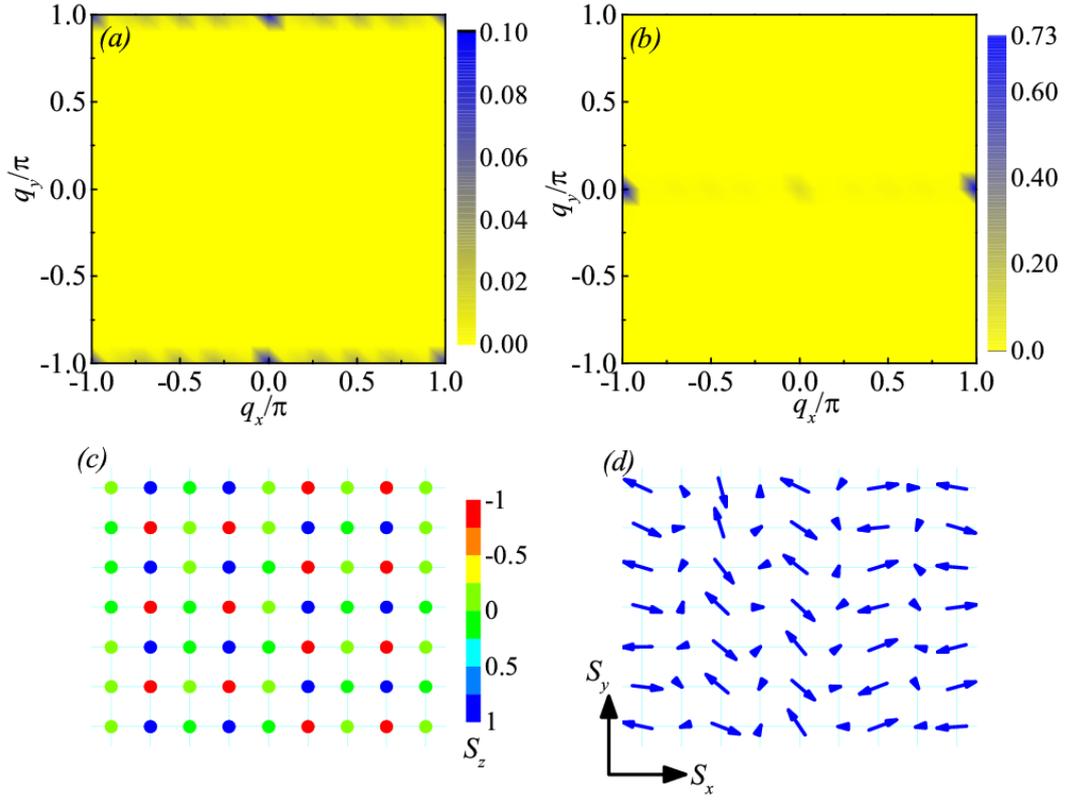

Figure 2. (color online) Distribution of the dipolar (a) and quadrupolar (b) magnetic structure factors at $T = 0.01$ for $J_1 = 0.1$, $K_1 = -0.5$, $J_O = 0$ and $\lambda = 0.6$. The $z$ component (c) and spin configuration on the $xy$-plane (d) of the AFQ order.

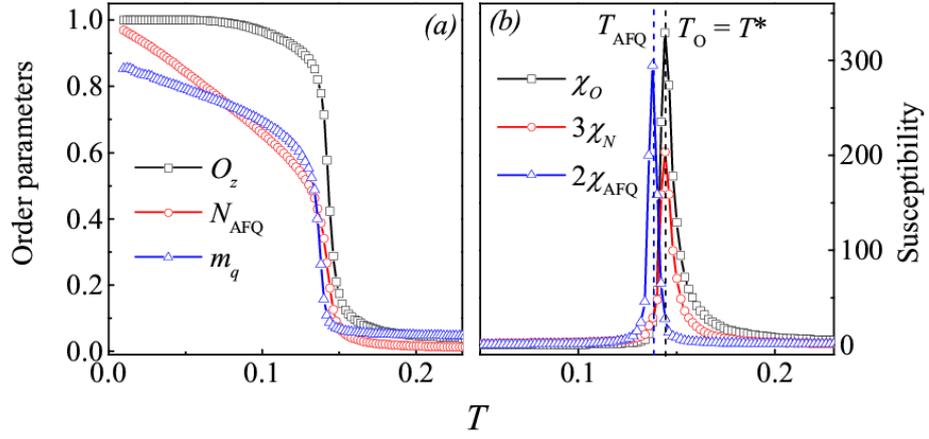

Figure 3. (color online) The calculated order parameters $O_Z$, $N_{AFQ}$, and $m_q$ (a) and their susceptibilities (b) as a function of $T$ for $J_1 = 0.1$, $K_1 = -0.5$, $J_O = 0$ and $\lambda = 0.6$.

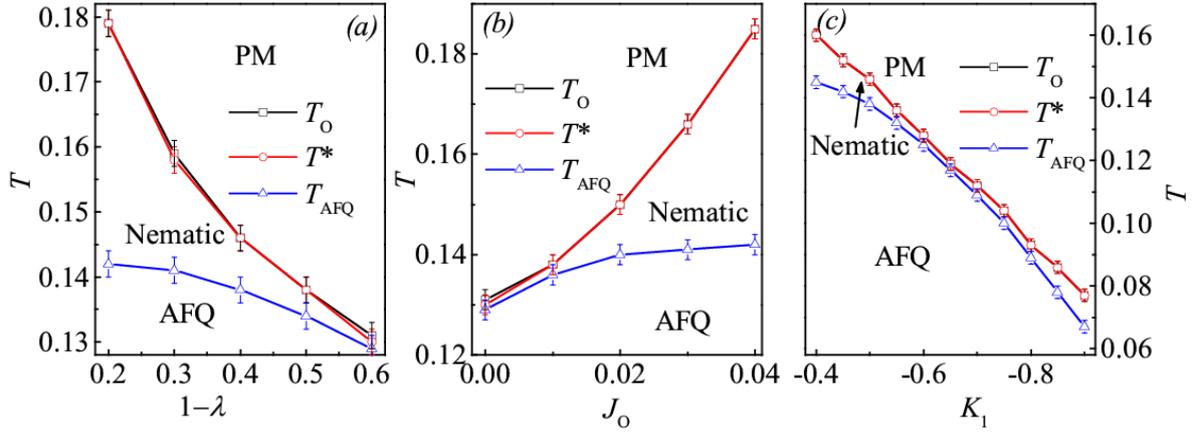

Figure 4. (color online) The simulated phase diagram (a) in the ($\lambda$, $T$) plane for $J_1 = 0.1$, $K_1 = -0.5$, and $J_O = 0$, (b) in the ($J_O$, $T$) plane for $J_1 = 0.1$, $K_1 = -0.5$, and $\lambda = 0.4$, and (c) in the ($K_1$, $T$) plane for $J_1 = 0.1$, $J_O = 0$, and $\lambda = 0.6$.

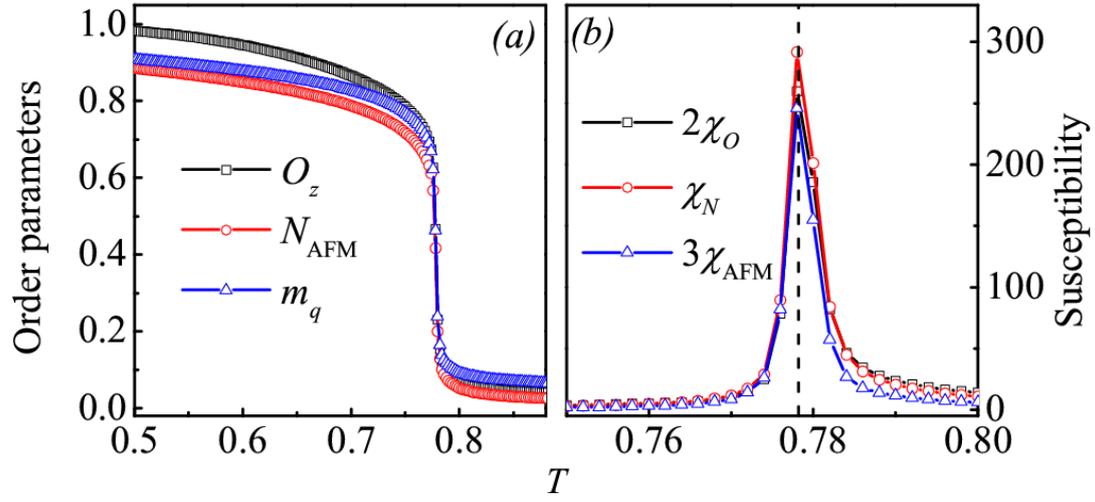

Figure 5. (color online) The calculated order parameters $O_Z$, $N_{AFM}$, and $m$ (a) and their susceptibilities (b) as a function of $T$ for $J_1 = 1.0$, $K_1 = 0.5$, $J_O = 0$ and $\lambda = 0.95$.

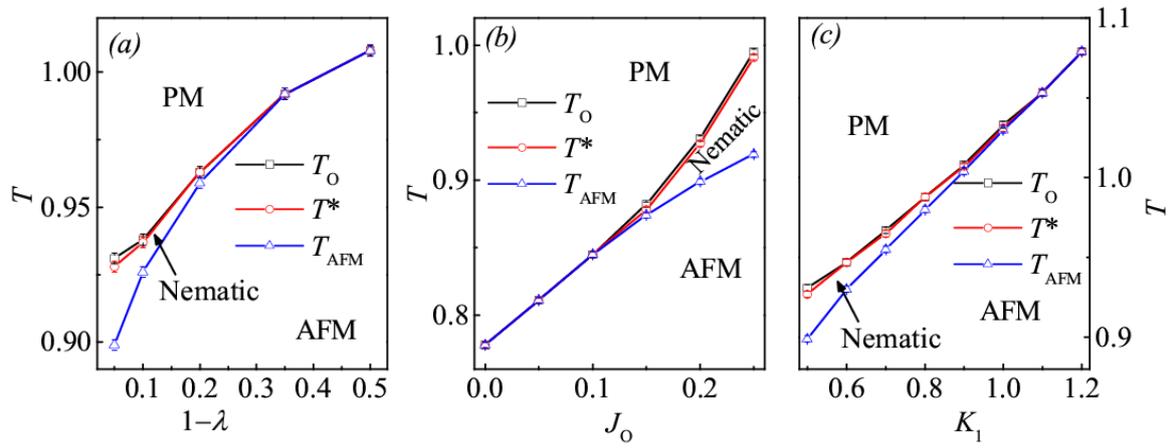

Figure 6. (color online) The simulated phase diagram (a) in the ($\lambda$, $T$) plane for $J_1 = 1.0$, $K_1 = 0.5$, and $J_O = 0.2$, and (b) in the ($J_O$, $T$) plane for $J_1 = 1.0$, $K_1 = 0.5$, and $\lambda = 0.95$, and (c) in the ($K_1$, $T$) plane for $J_1 = 1.0$, $\lambda = 0.95$, and $J_O = 0.2$.

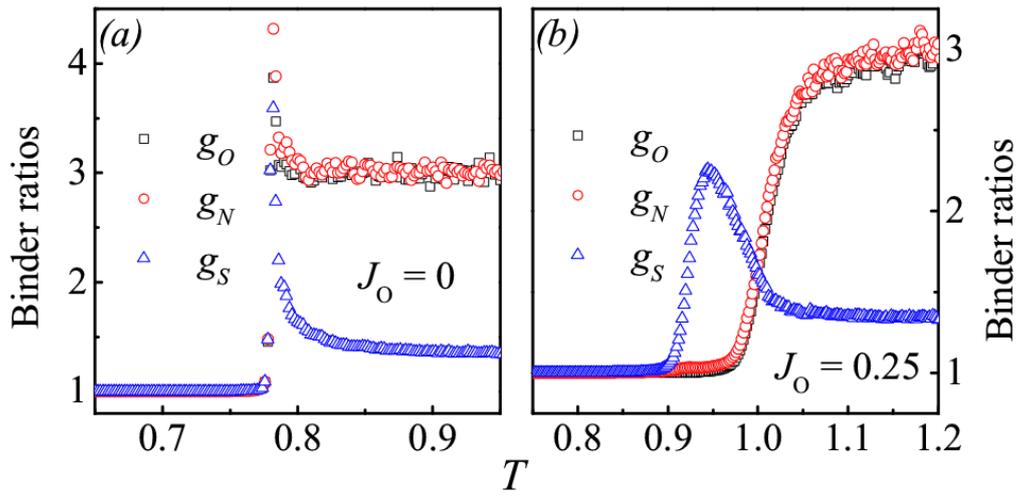

Figure 7. (color online) Calculated $g_O$, $g_N$, and $g_S$ as a function of $T$ for $J_1 = 1.0$, $K_1 = 0.5$, and $\lambda = 0.95$ at (a) $J_O = 0$, and (b) $J_O = 0.25$.